\documentclass[aps,prl,twocolumn,10pt]{revtex4-1}
\usepackage{mathrsfs}
\usepackage{bbm}
\usepackage{amsmath}
\usepackage{amssymb}
\usepackage{amsthm}
\usepackage{graphicx}
\usepackage{MnSymbol}
\usepackage{mathtools}
\usepackage[italicdiff]{physics}
\usepackage{chemformula}
\makeatletter
\usepackage{hyperref}
\usepackage{color}
\definecolor{supcol}{RGB}{10,50,180}
\definecolor{eqcol}{RGB}{220,10,100}
\hypersetup{colorlinks,citecolor=supcol,linkcolor=eqcol,urlcolor=supcol
}
\allowdisplaybreaks

\newcommand{\sectionprl}[1]{{\em #1}\/.---}

\begin{document}
\title{Brownian yet Non-Gaussian Heat Engine}
\author{I. Iyyappan}
\email[email: ]{iyyappan@imsc.res.in}
\affiliation{The Institute of Mathematical Sciences,\\ CIT Campus, Taramani, Chennai 600113, India.\\
Homi Bhabha National Institute, Training School Complex, Anushakti Nagar, Mumbai 400094, India.}
\author{Jetin E. Thomas}
\email[email: ]{jetinthomas@gmail.com}
\author{Sibasish Ghosh}
\email[email: ]{sibasish@imsc.res.in}
\affiliation{The Institute of Mathematical Sciences,\\ CIT Campus, Taramani, Chennai 600113, India.\\
Homi Bhabha National Institute, Training School Complex, Anushakti Nagar, Mumbai 400094, India.}

\date{\today}

\begin{abstract}
We investigate the performance of a Brownian heat engine working in a heterogeneous thermal bath where the mobility fluctuates. Brownian particle is trapped by the time-dependent harmonic potential, by changing the stiffness coefficient and the bath temperatures, we perform a Stirling cycle. We numerically evaluated the average work, power and efficiency. We compare our results with the Brownian heat engine working in a homogeneous thermal bath. We find that for the normal diffusive system, the performance of a Gaussian heat engine serves as an upper bound. We also observe that the non-Gaussian position distribution decreases the stochastic heat engine performance. 
\end{abstract}

\pacs{}
\maketitle
\sectionprl{Introduction}Exactly two hundred years ago, in 1824, Sadi Carnot discovered the universal upper bound for heat engine efficiency, $\eta_C=1-T_c/T_h$ \cite{carn}. Here, $T_h$ and $T_c$ are the temperatures of hot and cold reservoirs, respectively. The power output of the Carnot engine is zero, which has no use. Finite power is crucial for practical applications. Therefore, efficiency at maximum power (EMP) is another fundamental parameter to study real heat engines. Novikov for atomic power plants, Curzon and Ahlborn for endo-reversible heat engines showed that the EMP is $\eta_{NCA}=1-\sqrt{T_c/T_h}$ \cite{nov125, cur22}. EMP for heat engines has been actively investigated in the last four decades \cite{and208,and157,rub127,sal211,dev570,sal354,van190,esp150,joh052,tu312,ben230,izu100,bra070,her037,ang746,izu180,wan062,iyy500,rya050,pro220,iyy012,joh0121,joh044}.

Apart from macroscopic heat engines, micrometre-sized heat engines have unique features where the fluctuations become significant \cite{bus43}. Sekimoto identified the thermodynamic quantities, such as heat, work and internal energy for a Brownian particle at a single trajectory level \cite{sek123}. Now the framework is called stochastic thermodynamics \cite{seki,sei126}. Schmiedl and Seifert modelled a stochastic Carnot heat engine and showed that the EMP is $\eta_{SS}=\eta_C/(2- \alpha\eta_C)$ \cite{sch200}, where $\alpha$ represents the dissipation due to the finite-time processes. Using advanced experimental techniques, various microscopic heat engines have been realized in the lab \cite{bli143,qui588,mar67}. Subsequent studies on passive Brownian heat engines can be found Refs. \cite{hol050,ran042,tu052,pro041,arg052,hol120,sch068,nak012,gom643,miu042,ye043,miu034,che024,lin022}. However, Krishnamurthy \textit{et. al} showed that for Brownian heat engine working in a bacterial bath has a giant improvement in work output and efficiency \cite{kri113}. Hence active heat engines got special attention \cite{zak193,wul050,mar600,sah113,pie041,sah094,hol043,eke010,kum032,lee032,hol060,sza042,gro052,lee024,spe012,oh024}. 

Wang \textit{et. al} discovered a new class of diffusion process \cite{wan151,wan481}, it shows $\langle x^{2}(t)\rangle \propto t$ with the Laplace distribution of position (for a shorter time)
\begin{equation}\label{a}
	\rho(x,t)\backsimeq\mbox{exp}\left(\frac{-|x|}{\lambda(t)}\right).
\end{equation}  
Here, the characteristic decay length $\lambda(t)$ varies as $\sqrt{t}$. Many physical and biological systems exhibit normal diffusion with the non-Gaussian distribution of position \cite{hap111,lep198,ska256, gua333,mat042,cha022}. It is named a Brownian yet non-Gaussian diffusion (BNGD). This phenomenon can be well explained by the concept of super-statistics \cite{wan481,hap111} and the diffusing diffusivity (DD) model \cite{chu098,che021}. In the DD model, the diffusion coefficient is treated as a random variable and it is given by the square of the Ornstein–Uhlenbeck (OU) process \cite{che021}. 

In this Letter, we investigate the performance of Brownian heat engine working in a heterogeneous thermal bath and compare it with the performance of a Brownian heat engine working in a homogeneous thermal bath. Our results show that for the normal diffusive system, the performance of the Gaussian heat engine serves as an upper bound (see Figs. (\ref{fig:wpe})). The non-Gaussian distribution of position reduces the stochastic heat engine performance (see Figs. (\ref{fig:wpe}), (\ref{fig:ngp_x}), and (S4)). 

\sectionprl{Stochastic thermodynamics}We only focus on the overdamped regime. The following definitions are used to calculate the average thermodynamic quantities. The average internal energy is given by \cite{sch200}
\begin{equation}\label{1}
	U =\int V(x,\lambda (t)) \rho(x,t) dx.
\end{equation}
The average work done \textit{by} the particle is defined as \cite{seki}
\begin{equation}\label{2}
	W=\int_{t_i}^{t_f}dt \int \frac{\partial V(x,\lambda (t))}{\partial \lambda}\dot{\lambda} \rho(x,t) dx.
\end{equation}
Here, $V(x,\lambda (t))$ is the external potential and $\lambda (t)$ is the time-dependent control parameter. $\rho(x,t)$ is the probability distribution of position $x$ at time $t$. The justification for calling Eq. (\ref{2}) as work can be found in Refs. \cite{sek123} and \cite{jar329}. Using the first law like energy balance, we can find the absorbed heat from the thermal bath as \cite{ran042}
\begin{equation}\label{3}
	Q=W-\Delta U,
\end{equation}
where $\Delta U$ is the average change in internal energy during the isothermal process.

\sectionprl{The Model}When Brownian particles diffuse in the heterogeneous medium it encounter varying mobility with time. It can be due to the dynamical evolution of the medium or due to the heterogeneity of the medium \cite{hap111,lep198,ska256, gua333,mat042,cha022,kha064}. Chechkin \textit{et. al} used the set of Langevin equations to explain the diffusion of Brownian particles in an environment with fluctuating mobility \cite{che021}. Here, we consider the Brownian particle is trapped by the time-dependent harmonic potential. Therefore, the Langevin equation of motion is governed by the following equations 
\begin{equation}\label{f}
\frac{dx}{dt}=-\mu(t)\lambda(t)x+\sqrt{2\mu(t)\kappa_B T}\;\xi(t),
\end{equation}
\vspace{-0.7cm}
\begin{equation}\label{g}
\mu(t)=\chi^{2}(t),
\end{equation}
\vspace{-0.7cm}
\begin{equation}\label{h}
\frac{d\chi(t)}{dt}=-\frac{\chi(t)}{\tau'}+\sigma\zeta(t).
\end{equation}
Here, $x(t)$ is the position of the Brownian particle at time $t$. $\lambda(t)$ is the stiffness coefficient. We restrict our studies to one dimension. $\mu(t)$ is the fluctuating mobility \cite{che021}. $\kappa_B$ is the Boltzmann constant and $T$ is the bath temperature. $\xi(t)$ is the Gaussian white noise with $\langle\xi(t)\rangle=0$, and $\langle\xi(t)\xi(t')\rangle=\delta(t-t')$. $\mu(t)$ is defined as a square of the random variable $\chi$ to make sure it is positive. The random variable $\chi(t)$ is given by the OU process (Eq. (\ref{h})) and its explanation can be found in Ref. \cite{che021}. $\tau'$ is the correlation time of the OU process. $\zeta(t)$ is also a Gaussian white noise with $\langle\zeta(t)\rangle=0$, and $\langle\zeta(t)\zeta(t')\rangle=\delta(t-t')$. $\sigma$ is the strength of the fluctuating noise $\zeta(t)$. For simplicity, we consider the single random variable $\chi(t)$. In general, $\chi(t)$ can have $n$ degrees of freedom. The distribution of $\chi(t)$ at time $t$ is denoted by $f(\chi,t)$ and it evolves according to the Fokker-Planck equation \cite{risk}
\begin{equation}\label{i}
\frac{\partial f(\chi,t)}{\partial t}= -\frac{\partial }{\partial \chi} \left[-\frac{\chi}{\tau'} -\frac{\sigma^{2}}{2}\frac{\partial}{\partial \chi}\right]f(\chi,t).
\end{equation}
For a stationary state, the distribution of $\chi$ becomes 
\begin{equation}\label{j}
f(\chi)=\frac{1}{\sqrt{\pi\sigma^{2}\tau'}}\mbox{exp}\left(-\frac{\chi^{2}}{\sigma^{2}\tau'}\right).  
\end{equation}
We use the Euler-Muryama numerical method to integrate the Eqs. (\ref{f})-(\ref{h}). We set $x(0)=0$, and $\chi(0)$ has been chosen randomly from Eq. (\ref{j}). 

\begin{figure}[hpt]
	\centering
	\includegraphics[scale=0.56,angle=0]{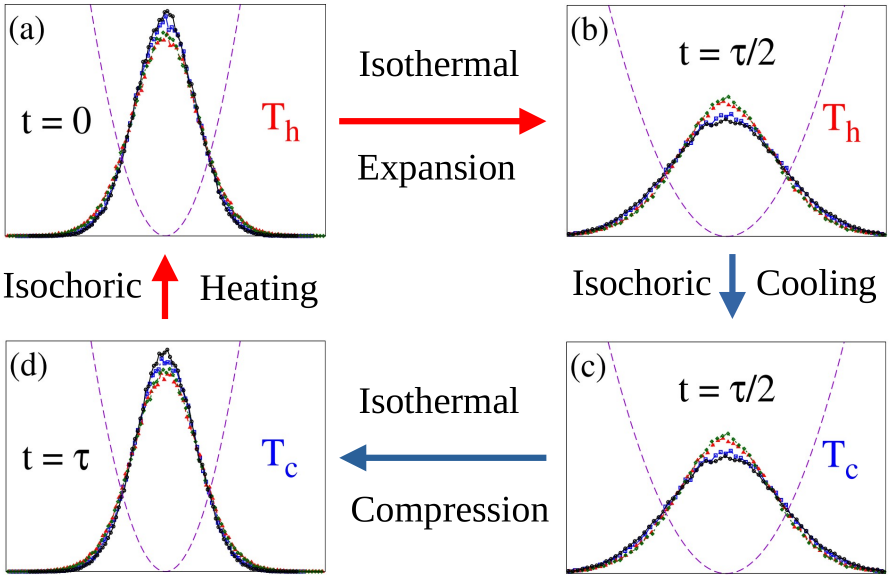}
	\vspace{-0.3cm}
\caption{\label{fig:fig0} The schematic diagram of the Brownian Stirling cycle. The dashed purple parabolic curve represents the harmonic potential. The probability distribution of $x$ for cycle time $\tau=0.1$ is depicted. The black solid line for a Brownian heat engine. The blue, red, and green curves for a Brownian yet non-Gaussian heat engine for the different sets of $\tau '$ and $\sigma$ obeying Eq. (\ref{amu}).}
\end{figure}

\begin{figure}
\includegraphics[scale=1.45]{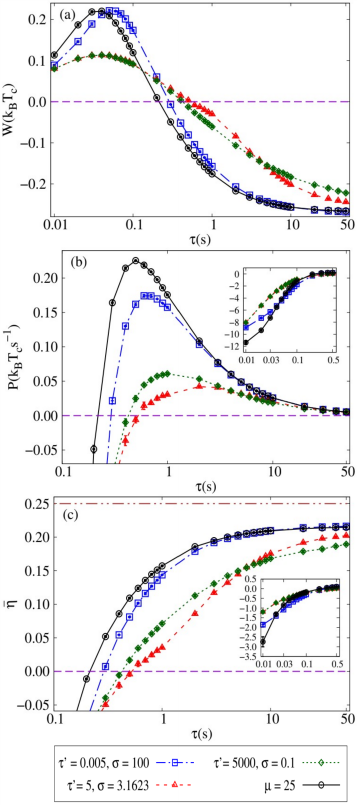}
\vspace{-0.6cm}
\caption{(a) The average work, (b) average power, and (c) average efficiency are plotted as a function of $\tau$ (logarithmic scale). The solid black line for BHE with constant mobility $\mu=25$. The blue, red, and green curves for BNGHE with different sets of $\tau'$ and $\sigma$ as indicated (with $\langle\mu \rangle=25$). Horizontal brown dot-dashed line in Fig. (c) represents the Carnot efficiency $\eta_C=0.25$. Insets are for shorter cycle times.}
\label{fig:wpe}
\end{figure}

Now, we construct a thermodynamic cycle to study its energetics. Experimentally realizing the Carnot cycle for Brownian particles are very difficult they contain adiabatic processes. However, Mart\'{i}nez \textit{et. al} accomplished a microscopic adiabatic process for an underdamped Brownian particle by precisely controlling the phase space volume \cite{mar120}. Nevertheless, the microscopic adiabatic process for a non-Gaussian system is not yet realized. Therefore, we study the Stirling cycle, which consists of two isothermal processes at different temperatures and two isochoric processes which connect them. The schematic diagram of our Brownian Stirling cycle is given in Fig. \ref{fig:fig0}.

\sectionprl{Isothermal expansion process}The Brownian particle is kept in a hot reservoir at temperature $T_h$ with the initial stiffness coefficient $\lambda_h$. We linearly decrease the stiffness coefficient ($\lambda_1(t)=\lambda_h-2\Delta\lambda t/\tau$, $\Delta\lambda\equiv \lambda_h-\lambda_l$ \cite{gom643}) for the period $t=0$ to $t=\tau/2$, which is easily realizable in experiments \cite{bli143}. This is equivalent to the volume expansion of the macroscopic heat engine in 3-dimension. We perform $m$ cycles before calculating the thermodynamic quantities \cite{supp}. The work done \textit{by} the particle  during the $(m+1)$th cycle is given by 
\begin{equation}\label{l}
	w_1=-\frac{\Delta\lambda}{\tau}\int_{m\tau}^{(m+1/2)\tau} x^{2}(t)dt.
\end{equation}
The change in internal energy is given by $\Delta u_1=\left[\lambda_l x^{2}([m+1/2]\tau)- \lambda_h x^{2}(m\tau)\right]/2$. Using Eq. (\ref{3}), we can calculate the heat absorbed during the isothermal expansion process as $q_h=w_1-\Delta u_1$. 

\sectionprl{Isochoric Cooling}Keeping the stiffness coefficient constant, we decrease the temperature of the bath from $T_h$ to $T_c$ instantaneously \cite{exp}. Therefore, $\rho(x,t)$ does not evolve. The work done by the particle during an isochoric process is zero since there is no change in the external control parameter \cite{bli143}.

\sectionprl{Isothermal compression process}Now, keeping the bath temperature at $T_c$, we linearly increase the stiffness coefficient   ($\lambda_3(t)=\lambda_l-\Delta\lambda[1-2t/\tau]$) for the time period $t=\tau/2$ to $t=\tau$ \cite{gom643}. The work done \textit{on} the particle is 
\begin{equation}\label{n}
	w_3=\frac{\Delta\lambda}{\tau}\int_{(m+1/2)\tau}^{(m+1)\tau} x^{2}(t)dt.
\end{equation}
The change in internal energy during the isothermal compression process is $\Delta u_3=\left[\lambda_h x^{2}([m+1]\tau)-\lambda_l x^{2}([m+1/2]\tau) \right]/2$. Therefore, the heat ejected to the cold reservoir becomes $q_c=w_3-\Delta u_3$.

\sectionprl{Isochoric Heating}Finally, the bath temperature increased from $T_c$ to $T_h$, instantaneously, while keeping the stiffness coefficient constant \cite{exp}. Again, the isochoric heating process does not contribute to the work. The total work output delivered by the Brownian engine is $-w=-w_1+w_3$ which is random. Therefore, the average work output is given by $-W=-\langle w_1\rangle+\langle w_3\rangle$. 

\sectionprl{Numerical Simulation}We consider the ensembles of $10^{5}$ particle trajectories to compute the average quantities. We set $dt=10^{-5}s$. The high and low values of stiffness coefficients, respectively as $\lambda_h=5\; p\mbox{N}/\mu m$ (pN-pico Newton), and $\lambda_l=1\; p\mbox{N}/\mu m$ \cite{gom643}. Temperatures of the hot and cold reservoirs, respectively as $T_h=400$ K, and $T_c=300$ K. $\kappa_B=1.38\times10^{-5}\; p\mbox{N}\mu m\mbox{K}^{-1}$. The average mobility becomes \cite{supp} 
\begin{equation}\label{amu}
	\langle\mu \rangle=\frac{\sigma^{2}\tau '}{2}.
\end{equation}
We study the three different sets of $\sigma$ and $\tau '$ which satisfy $\langle\mu \rangle=25\mu m pN^{-1}s^{-1}$. We show this set of values satisfying the fluctuation-dissipation theorem \cite{supp}. To compare our results with the homogeneous thermal bath, we set the mobility $\mu=25\mu m pN^{-1}s^{-1}$ and consider the following Langevin equation
\begin{equation}\label{}
	\frac{dx}{dt}=-\mu\lambda(t)x+\sqrt{2\mu\kappa_B T}\varsigma(t),
\end{equation}
$\varsigma(t)$ is the Gaussian white noise with $\langle\varsigma(t)\rangle=0$, and $\langle\varsigma(t) \varsigma(t') \rangle= \delta(t-t')$.

\sectionprl{Main Results}Our findings are presented in this section. All figures are plotted in a semi-logarithmic coordinate (either x-axis or y-axis) except Fig. (\ref{fig:ngp_x}). Throughout this Letter, the black solid line represents the Brownian heat engine (BHE). The blue, green, and red curves are representing the Brownian yet non-Gaussian heat engine (BNGHE). The mean work output is plotted as a function of cycle time $\tau$ in Fig. (\ref{fig:wpe}a).  For initial cycle times, the engine consumes work and it starts to deliver work after a certain value of $\tau$. We find that work output saturates at larger cycle times. This work behavior (from $\tau=0.1$ to $\tau=50$) can be well fitted by using the low-dissipation model \cite{sch200,esp150}, where the work is given by \cite{bli143}
\begin{equation}\label{wk}
-W=-W_{\mbox{qs}}+\frac{\Sigma}{\tau}.
\end{equation}
Here, $W_{\mbox{qs}}$ is the quasi-static work (see Ref. \cite{bli143} for its expression), and $\Sigma/\tau$ is the dissipated work. $\Sigma$ contains the information about the protocol, and system bath interaction. The value of $\Sigma$ is different for each case. BHE delivers the highest work as compared with the BNGHE. However, BNGHE with $\tau '=0.005$, and $\sigma=100$ (blue curve) merges with BHE around $\tau=10$, and the other two cases of BNGHE (red, and green curves) approaches the BHE when the non-Gaussian parameter for the position (Eq. (\ref{ngp})) approaches zero (see Figs. (\ref{fig:ngp_x}), and (S5)). Now, we analyze the physical reason behind the higher work output of BHE as compared with the BNGHE. For that, we plotted the $\langle x(t)^{2}\rangle$ during the cycle times in Fig. (S3) \cite{supp}. It shows that $\langle x(t)^{2}\rangle$ for a BHE covers a larger area than the BNGHE which directly gives rise to the higher work (see Eq. (\ref{2}), and Fig. (S3)). Nevertheless, this difference in the area becomes smaller and smaller when $\tau$ turns larger and larger.

The performance of a stochastic heat engine is given by the average power and efficiency
\begin{equation}
P=\frac{-W}{\tau},\;\;\;\;\bar{\eta}=\frac{-W}{Q_h}.
\end{equation}
Fig. (\ref{fig:wpe}b) shows that the average power becomes positive after a particular cycle time and it increases with $\tau$. The power attains its maximum ($P_{max}$) at $\tau^{\star}=2\Sigma/W_{qs}$ due to its functional form given in Eq. (\ref{wk}). Then starts to decrease monotonically with $\tau$. Since the rate at which the $-W$ increases with cycle time decreases for a larger $\tau$.  We find that $P_{max}\mbox{(BHE)}> P_{max}\mbox{(BNGHE)}$. The power asymptotically approaches zero when $\tau$ is very large.

The average efficiency is plotted as a function of $\tau$ in Fig. (\ref{fig:wpe}c). The $\bar{\eta}$ increases monotonically with the cycle time. We find that the $\bar{\eta}$ (BHE) and $\bar{\eta}$ (BNGHE) blue curves are merging at larger cycle times. However, for intermediate cycle times, the $\bar{\eta}$ (BHE) is always higher than the $\bar{\eta}$ (BNGHE). One can notice that $W$, $P$, and $\bar{\eta}$ for the blue curves are slightly higher than the black curve at larger cycle times. This is due to the smaller number of trajectories for the DD system (see S4 of \cite{supp} for the justification). Therefore, for the normal diffusion, Brownian heat engine with Gaussian position distribution has the maximum performance. 

\begin{figure}
	\includegraphics[scale=0.6]{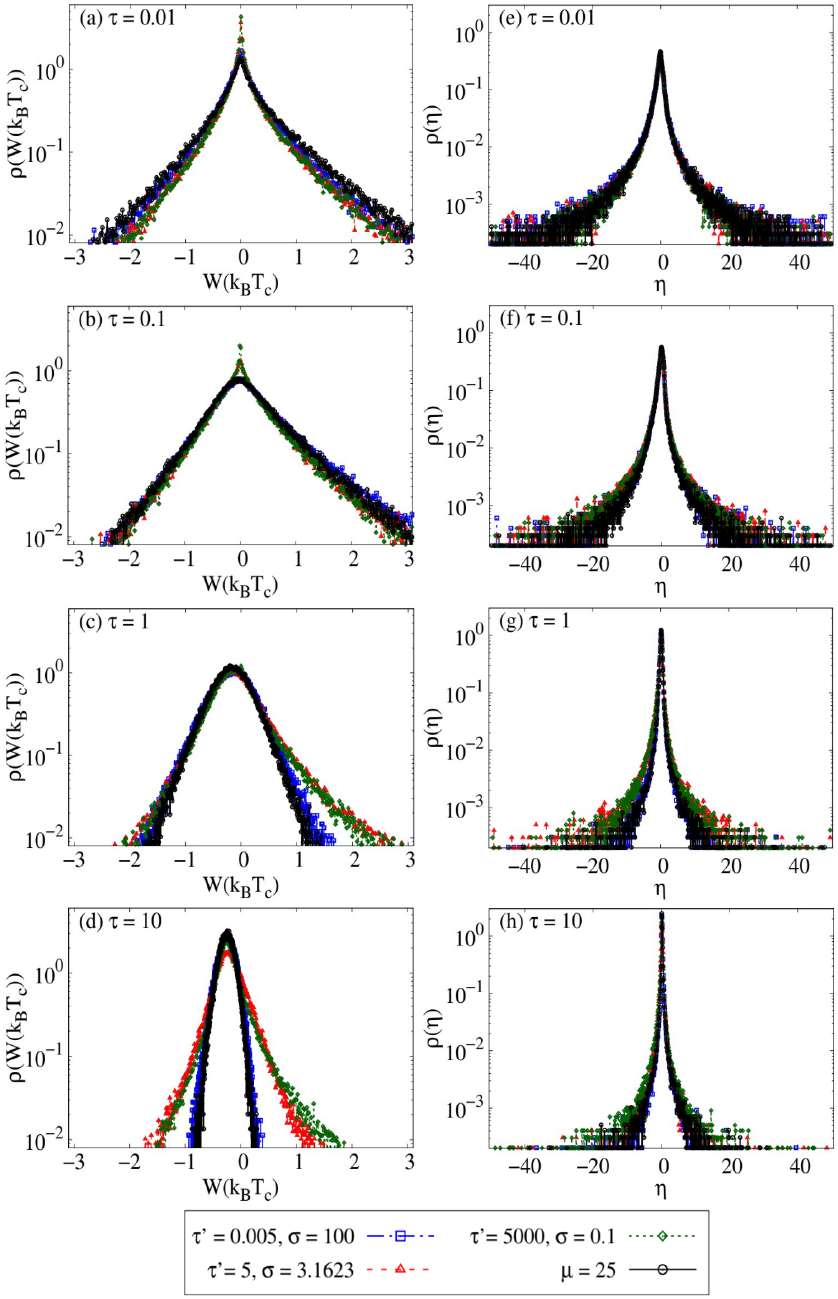}
	\vspace{-0.3cm}
	\caption{(a-d) The distribution of work, and (e-h) the distribution of efficiency are plotted for different cycle times as mentioned. The black line for BHE with $\mu=25$. The blue, red, and green curves for BNGHE with different sets of $\tau'$ and $\sigma$ as indicated (with $\langle\mu \rangle=25$). Y-axis in logarithmic scale.}
	\label{fig:pdwe}
\end{figure}

\begin{figure}
\includegraphics[scale=0.625]{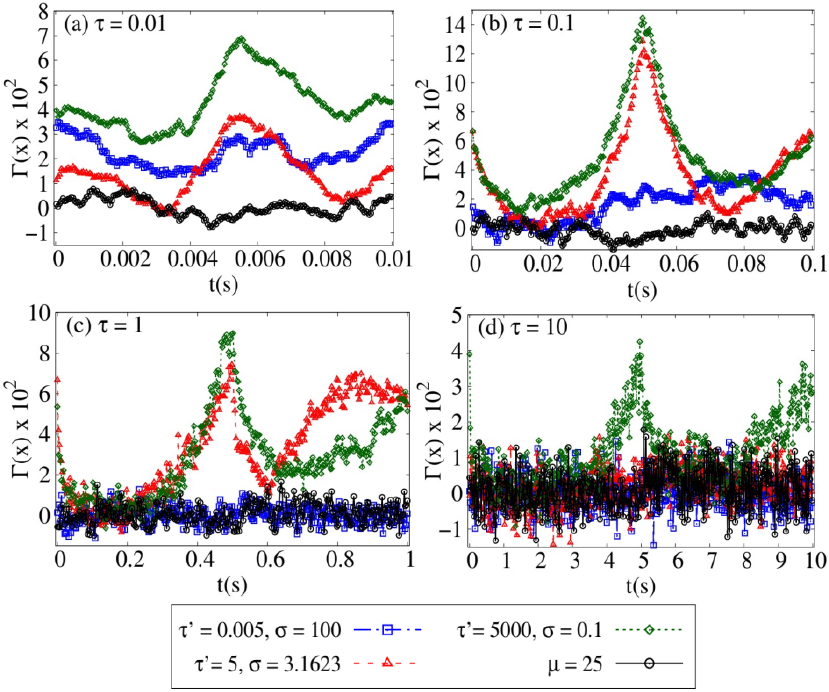}
\vspace{-0.5cm}
\caption{The non-Gaussian parameter is plotted as a function of $t$ at cycle times, (a) $\tau=0.01$, (b) $\tau=0.1$, (c) $\tau=1$, and (d) $\tau=10$. The solid black line for BHE with constant mobility $\mu=25$. The blue, red, and green curves for BNGHE with different sets of $\tau'$ and $\sigma$ as indicated (with $\langle\mu \rangle=25$).}
\label{fig:ngp_x}
\end{figure}

\sectionprl{The distribution of work and efficiency} The work, and efficiency distributions are plotted at cycle times $\tau=0.01,\;0.1,\;1,\; \mbox{and}\;10$ in Fig. (\ref{fig:pdwe}). Our results show that for a shorter cycle time, the work of BHE spreads out much more than the BNGHE. However, when $\tau$ increases the spread interchanges. For $\tau=10$, the spread of BHE is smaller than the BNGHE. Additionally, BNGHE with $\tau '=5000$ and $\sigma=0.1$ (green curve) has the greater spread compared with the other two cases of BNGHE (red and blue) at the larger $\tau$. The efficiency distribution of BHE and BNGHE are similar for a shorter time. However, at a larger time, the spread of efficiency for BHE is smaller as compared with BNGHE.

\sectionprl{Non-Gaussian parameter} To find the effect of non-Gaussian distribution of position on the performance of the BHE, we plotted the $\Gamma(x)$ during the cycle times at $\tau=0.01,\;0.1,\;1,\; \mbox{and}\;10$ in Fig. (\ref{fig:ngp_x}). The non-Gaussian parameter for the position is given by \cite{che252}
\begin{equation}\label{ngp}
\Gamma(x)=\frac{k(x)}{3}-1.
\end{equation}
Here, the \textit{kurtosis}, $k(x)=\langle (x-\langle x\rangle)^{4} \rangle/\mbox{Var}(x)^{2}$ \cite{raha40}, and the \textit{variance}, $\mbox{Var}(x)=\langle (x-\langle x\rangle)^{2} \rangle$. Fig. (\ref{fig:ngp_x}) shows that the red and green curves have a higher $\Gamma(x)$ than the black and blue curves which has a direct consequence on the performance of heat engines (see Fig. (\ref{fig:wpe})). The higher $\Gamma(x)$ reduces the performance of stochastic heat engine (see red, and green curves Figs. (\ref{fig:wpe}), (\ref{fig:ngp_x}), and (S4)). For cycle time $\tau=10$ (Fig. (\ref{fig:ngp_x}d)), we find that the $\Gamma(x)$ of the red curve reduced drastically as compared with $\tau=1$ (see Fig. (\ref{fig:ngp_x}c)) which is also reflected in the performance of the heat engine (red curve surpassing the green curve) (see Fig. (\ref{fig:wpe})).

\sectionprl{Conclusion}In this Letter, we studied the performance of a Brownian yet non-Gaussian heat engine in detail, where the inhomogeneity of the thermal bath is considered and we have compared our results with the Brownian heat engine working in a homogeneous thermal bath. We find that for a particular set of $\tau '$ and $\sigma$, the BNGHE performance merges with the BHE performance blue and black curves, respectively (see Figs. (\ref{fig:wpe}), (\ref{fig:ngp_x}), and (S5)) when $\Gamma(x)$ of BNGHE approaches $0$. Further, the BNGHE performance decreased by the presence of non-Gaussianity in the system. 

For a BHE working in a bacterial bath (which is super-diffusive), it is shown that the increase in non-Gaussian position distribution enhances the efficiency see Fig. (3) of Ref. \cite{kri113} which is in contrast to our results (normal diffusion). However, Datta \textit{et. al} pointed out that for a non-equilibrium thermal bath, the second law of thermodynamics needs to be modified \cite{dat031}. Therefore, Whether the non-Gaussian position distribution increase or decrease the performance of BHE for the super-diffusive system needs to be analyzed. While writing our Letter, we noticed the work of Sposini \textit{et. al} \cite{spo117}. They showed that the non-Gaussian diffusion is useful when only a few searchers are enough to reach the target like human fertilization and when a large of searchers need to find the target non-Gaussian distribution has a disadvantage.  

\sectionprl{Future outlook}The Brownian yet non-Gaussian heat engine can be realized in the lab. Spatial heterogeneity can be created artificially \cite{cha022} or one can use the Entangled F-actin networks \cite{wan151}. With the optical tweezers, Brownian particle could be trapped by harmonic potential. Changing the laser intensity and bath temperatures, the Stirling cycle can be accomplished \cite{bli143}. It will be interesting to study the energetics of DD models for active systems \cite{kha064}. Further, in our future study, we would like to investigate the BNGHE with the external fluctuating force
\cite{mar032}, and also interested in extending the present work to the quantum domain. For an anomalous (sub or supper) diffusive system \cite{met339}, finding the correlation between the performance of BHE and the non-Gaussian distribution of position would be intriguing. It is challenging to prove our results analytically.

\begin{acknowledgments}
S. G. acknowledges the support from Interdisciplinary Cyber Physical Systems (ICPS) program of the Department of Science and Technology (DST), India, Grant No. DST/ICPS/QuEST/Theme-1/2019/13. I. I. sincerely thank K. Sekimoto, A. Dechant, K. Brandner, P. Pietzonka, and Y. Baek for their crucial comments and suggestions during the YITP-YSF Symposium on Perspectives on Non-Equilibrium Statistical Mechanics: The 45th Anniversary Symposium of Yamada Science Foundation (YITP-W-23-06).  J. E. T. thanks Ralf Metzler for the useful discussion. I. I. also thank Joseph Prabagar for his unwavering support and encouragement.

I. I. is extremely delighted to dedicate this work to the late Prof. S. V. M. Satyanarayana who taught free physics classes on Sundays to research aspirants for over 25 years. I. I is one of the beneficiaries of his \textit{Sunday class}.
\end{acknowledgments}

I. I. and J. E. T. contributed equally to this work.


\begin{thebibliography}{100}
\bibitem{exp}
In experiments, isochoric cooling (heating) takes a few milliseconds to cool (heat) the bath temperature from $T_h$ to $T_c$ ($T_c$ to $T_h$) \cite{bli143}, which can be neglected as compared with the cycle time.
\end{thebibliography}
\end{document}


\title{Supplemental Material: Brownian yet Non-Gaussian Heat Engine}
\author{I. Iyyappan}
\email[email: ]{iyyappan@imsc.res.in}
\affiliation{The Institute of Mathematical Sciences,\\ CIT Campus, Taramani, Chennai 600113, India.\\
Homi Bhabha National Institute, Training School Complex, Anushakti Nagar, Mumbai 400094, India.}
\author{Jetin E. Thomas}
\email[email: ]{jetinthomas@gmail.com}
\author{Sibasish Ghosh}
\email[email: ]{sibasish@imsc.res.in}
\affiliation{The Institute of Mathematical Sciences,\\ CIT Campus, Taramani, Chennai 600113, India.\\
Homi Bhabha National Institute, Training School Complex, Anushakti Nagar, Mumbai 400094, India.}

\begin{abstract}
In this supplemental material, we gave the details of the analytical and numerical results of our study. The derivation for average mobility, and the definition for time-periodic steady-state. The plots for average thermodynamic quantities such as input heat, ejected heat, and change in internal energy. We plotted the power and efficiency with error bars. Total non-Gaussian parameter per cycle time. Finally, we also presented our results for the free Brownian and Brownian yet non-Gaussian diffusion.
\end{abstract}

\pacs{}
\maketitle
\tableofcontents
\section{The average mobility of the fluctuation mobility medium}
The Fokker-Planck equation for the Langevin Eq. (7) becomes
\begin{equation}\label{i}
\frac{\partial f(\chi,t)}{\partial t}= -\frac{\partial }{\partial \chi} \left[-\frac{\chi}{\tau'} -\frac{\sigma^{2}}{2}\frac{\partial}{\partial \chi}\right]f(\chi,t). \tag{S2}
\end{equation}
Here, $f(\chi,t)$ is the probability distribution of $\chi$ at time $t$. Solving the Eq. (\ref{i}) with the initial value $\chi(0)\equiv\chi_0$, we get the following probability distribution 
\begin{equation}\label{k}
f(\chi,t|\chi_0,0)=\frac{1}{\sqrt{\pi \sigma^{2}\tau '\left(1-\exp\left[\frac{-2t}{\tau '}\right]\right)}}\exp\left[\frac{-\left(\chi-\chi_0\exp\left[-\frac{t}{\tau '}\right]\right)^{2}}{\sigma^{2}\tau '\left(1-\exp\left[\frac{-2t}{\tau '}\right]\right)}\right] . \tag{S3}
\end{equation}
Now, the $\chi_0$ is drawn randomly from the below steady-state  distribution 
\begin{equation}\label{j}
f(\chi_0)=\frac{1}{\sqrt{\pi\sigma^{2}\tau'}}\mbox{exp}\left(-\frac{\chi_0^{2}}{\sigma^{2}\tau'}\right).  \tag{S4}
\end{equation}
Taking the average over $\chi_0$ \cite{besL01}
\begin{equation}
f(\chi,t)=\int_{-\infty}^{\infty}f(\chi,t|\chi_0,0)f(\chi_0)d\chi_0. \tag{S5}
\end{equation}
We get the following probability distribution
\begin{equation}\label{prob}
f(\chi)=\frac{1}{\sqrt{\pi\sigma^{2}\tau'}}\mbox{exp}\left(-\frac{\chi^{2}}{\sigma^{2}\tau'}\right). \tag{S6}
\end{equation}
Using Eq. (\ref{prob}), we get the average mobility as
\begin{equation}\label{s5}
\langle \mu\rangle=\langle \chi^{2}\rangle=\frac{1}{2}\sigma^{2}\tau'. \tag{S7}
\end{equation}

\section{Time-periodic steady-state}
When we perform a Stirling cycle for a given cycle time $\tau$ to an ensemble of Brownian particles. The final position $x(\tau)$ (with $x(0)=0$) is a random variable. To get the efficiency independent of $x(0)$, first, we need to run many cycles until $\langle x(n\tau)^{2}\rangle$ and $\langle x([n+1]\tau)^{2}\rangle$ becomes equal, where $n$ is the number of cycles performed. After this, the probability distribution satisfies the following condition, $p_{ss}(x,t)=p_{ss}(x, t+\tau)$, which is called the time-periodic steady-state (TPSS) \cite{ran042,gom643,kum032}. If the ensemble of Brownian particles once reaches a TPSS, then it will no longer have the memory of its initial position $x(0)=0\,\mu m$(micrometer). Brownian particles in a homogeneous medium reach a steady state after a few cycles. However, we don't know in the beginning, the number of cycles required to reach the steady state. Therefore, we keep track of the $x^{2}$ at the end of each cycle and compare this with that of the previous cycle. We keep the precision 
\begin{equation}
\Omega\equiv \langle x([n+1]\tau)^{2} \rangle- \langle x(n\tau)^{2} \rangle < 10^{-6}.	
\end{equation}
Once the above condition is satisfied, we start to calculate the average quantities like work, heat supplied, and ejected heat for the cycles after this difference reaches a lower value than the fixed precision value. For cycles after this, the $x(m\tau)^{2}$ and the other average quantities at the end of each cycle behave like random variables fluctuating about some average values. We consider these values obtained at the end of cycles after reaching steady-state as identically distributed random variables. Consequently, the random variable of the average of the cumulative sum of the $x^{2}$ of each particle at the end of cycles after reaching steady-state obeys the central limit theorem. Thus, the standard deviation of this random variable for the $N$ particles decreases for each added cycle after reaching the steady state. We fixed another precision of $0.0007$ for this standard deviation which will enable us to calculate the average quantities with the desired statistical accuracy. Hence, the definition of the precision that determines the statistical accuracy from the central limit theorem is
\begin{equation}
\sqrt{\frac{1}{N}\sum_{i}^{N}\left(\frac{\sum_{m=1}^{N_{cycst}}x_{i}(m\tau)^{2}}{N_{cycst}}-\frac{\sum_{i=1}^{N}\sum_{m=1}^{N_{cycst}}x_{i}(m\tau)^{2}}{N*N_{cycst}}\right)^{2}} < 0.0007.	
\end{equation} 
where $N_{cycst}$ is the number of cycles traversed after reaching the steady-state and $m=1$ is the first cycle after reaching the steady state. We set $\Omega<10^{-7}$ for cycle times $0.01$ to $0.1$, and $\Omega<10^{-6}$ for cycle times $0.2$ to $10$. However, due to the limitation of our computational facility, we have used $\Omega<10^{-5}$ for cycle times $20$ to $50$. 
 
\section{The average input heat, ejected heat, and change in internal energy}
\begin{figure}[hpt]
\centering
\includegraphics[scale=1.45,angle=0]{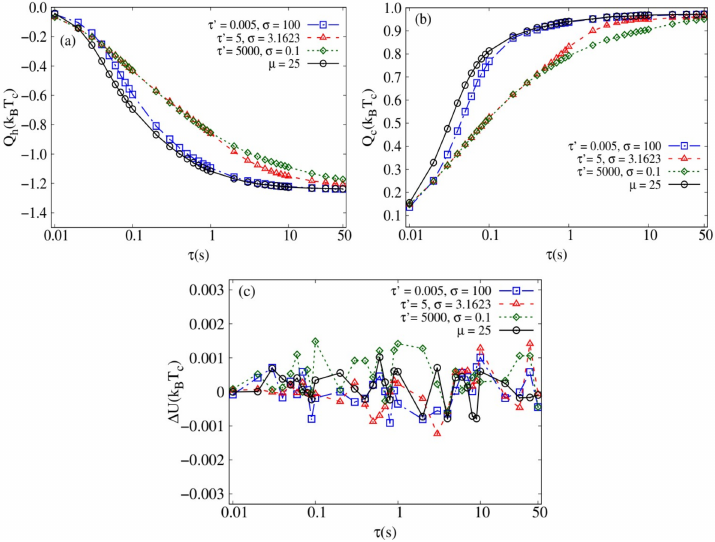}
\vspace{-0.4cm}
\caption{\label{fig:hci} (a) The average input heat, (b) average ejected heat, and (c) average change in internal energy are plotted as a function of cycle time $\tau$ (logarithmic scale). The solid black line for BHE with  $\mu=25$. The blue, red, and green curves for BNGHE with different sets of $\tau'$ and $\sigma$ as indicated obeying Eq. (12), $\langle\mu \rangle=\sigma^{2}\tau '/2=25$.}
\end{figure}

The average input heat is plotted in Fig. (\ref{fig:hci}a). It shows that $Q_h$ decreases monotonically with $\tau$ and saturates at a larger cycle time. We find that the Brownian heat engine (BHE) absorbs the larger input heat as compared with the Brownian yet non-Gaussian heat engine (BNGHE). The average ejected heat is plotted in Fig. (\ref{fig:hci}b). $Q_c$ increases monotonically with cycle time and saturates at a larger $\tau$. Again, the BHE ejects the larger heat as compared with the BNGHE. The average change in internal energy after completing a cycle is plotted as a function of $\tau$ in Fig. (\ref{fig:hci}c). It shows that the mean change in internal energy fluctuates around zero and it has to be noted that this fluctuation is a hundred times smaller than that of $Q_h$ and $Q_c$ values. The fluctuation in $\Delta U$ is due to the finite value of $dt=10^{-5}$ in our numerical integration of Eqs. (5) -(7) and (13) as well as due to the less number of trajectories (in our case $10^{5}$). See also Fig. (3) in Ref. \cite{bli070} for further explanations.  

\section{Power, and efficiency with error bars}
\begin{figure}[hpt]
	\centering
	\includegraphics[scale=1.5,angle=0]{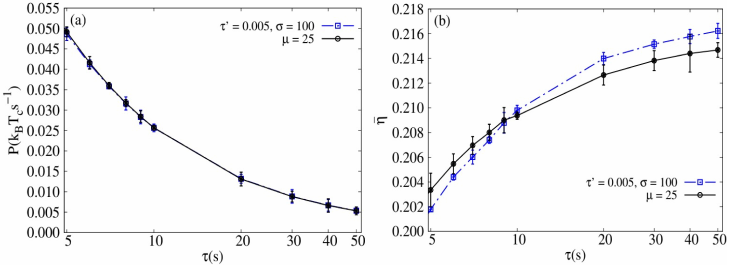}
	\vspace{-0.5cm}
	\caption{\label{fig:pe_eb} (a) The average power,  and (b) efficiency with error bars are plotted as a function of cycle time $\tau$ (logarithmic scale). The solid black line for BHE with $\mu=25$. The blue dot-dashed curve for BNGHE with $\tau'=0.005$, $\sigma=100$, and $\langle\mu \rangle=25$.}
\end{figure}

We plotted the power and efficiency with error bars for a cycle time $\tau=5$ to $\tau=50$ (so that the error bars are visible) and we only considered black (BHE) and blue (BNGHE) curves in Fig. (\ref{fig:pe_eb}). We observe that the BNGHE efficiency becomes higher than the BHE, it is because of numerical error, not a physical phenomenon. Our reasoning as follows, diffusing diffusivity system requires (at least) $10^{6}$ trajectories to minimize the fluctuations at a reasonable level (see Figs. \ref{fig:msd} (c), and  (d) for the difference) which is unfeasible for heat engine with our computational facility (where only $10^{5}$ trajectories are considered). However, Chechkin \textit{et. al} \cite{che021} showed that the diffusing diffusivity system approaches the Gaussian distribution of position at a very large time (see Figs. (1) and (3) of Ref. \cite{che021}). Although $10^{6}$ trajectories are considered in Ref. \cite{che021}, we could still see the fluctuation in \textit{Kurtosis} (see Figs. (3) of Ref. \cite{che021}, and also Figs. \ref{fig:msd} (c), and (d) for our case). The average work, and input heat depend on the $\langle x(t)^{2}\rangle$ through Eqs. (3) and (4). When Brownian yet non-Gaussian diffusion and Brownian diffusion have the same Gaussian distribution of position at a larger cycle time will lead only to the same power and efficiency. Therefore, the difference visible between the BHE and BNGHE performance in Fig. (\ref{fig:pe_eb}) is solely a numerical error.  

\section{The average position square}
\begin{figure}[hpt]
\centering
\includegraphics[scale=1.4,angle=0]{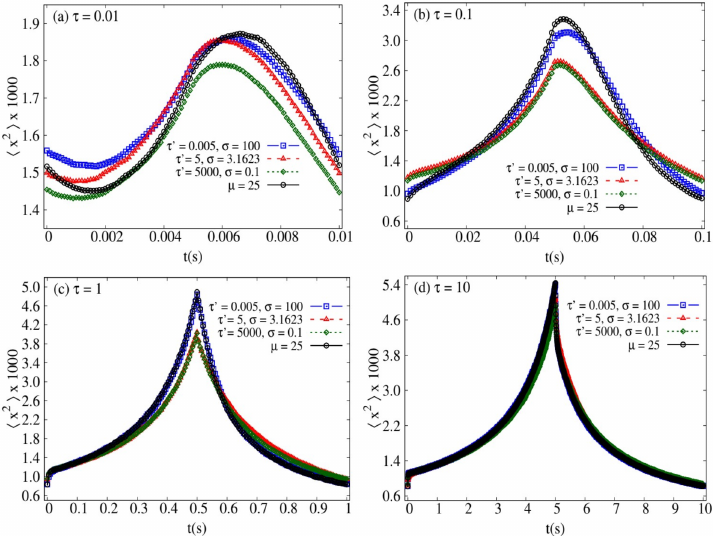}
\vspace{-0.4cm}
\caption{\label{fig:x2} The average $x^{2}$ is plotted as a function of time during the cycle times at (a) $\tau=0.01$, (b) $\tau=0.1$, (c) $\tau=1$, and (d) $\tau=10$. The solid black line for BHE. The blue, red, and green curves for BNGHE with different sets of $\tau'$ and $\sigma$ as indicated.}
\end{figure}

To understand the BHE and BNGHE work output, we plotted the $\langle x(t)^{2} \rangle$ during the cycle times at $\tau=0.01,\;0.1,\;1,\; \mbox{and}\;10$ in Fig. (\ref{fig:x2}). We find that for a cycle time, $\tau=0.01$, the $\langle x(t)^{2} \rangle$ keeps increasing even after the isothermal compression process starts (see Fig. \ref{fig:x2}a) which makes the total work positive. As we know work is the path variable, we get different work for different cases. Fig. (\ref{fig:x2} c, and d) shows that the black, and blue curves merge which gives rise to roughly equal work outputs. The asymmetry of $\langle x(t)^{2} \rangle$ which we (particularly) observe in Figs. (\ref{fig:x2}c) and (\ref{fig:x2}d) are the reasons behind the negative work. It has to be noted that the area under the curve is directly proportional to the average work (see Eq. (3)) with the -ve sign for time $t=0$ to $t=\tau/2$ and +ve for $t=\tau/2$ to $t=\tau$.

\section{Total non-Gaussian parameter per cycle time}
\begin{figure}[hpt]
	\centering
	\includegraphics[scale=1.2,angle=0]{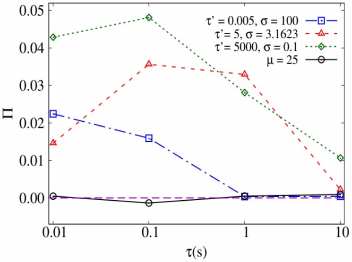}
	\vspace{-0.4cm}
	\caption{\label{fig:ingp} Total non-Gaussian parameter per cycle time is plotted as a function of cycle time. The solid black line for BHE with $\mu=25$. The blue, red, and green curves for BNGHE with different sets of $\tau'$ and $\sigma$ as indicated with $\langle\mu \rangle=25$. Horizontal dashed purple line for $\Pi(x)=0$.}
\end{figure}

To find the effect of non-Gaussian position distribution on the performance of a stochastic heat engine, we calculate the total non-Gaussian parameter per cycle time using the following equation
\begin{equation}\label{ingp}
\Pi(x)=\frac{\int_{0}^{\tau}\Gamma(x,t)dt}{\tau}.
\end{equation}
We plotted Eq. (\ref{ingp}) in Fig. (\ref{fig:ingp}). We find that for heat engine at $\tau=1$, and $\tau=10$ the higher $\Pi(x)$ decreases the stochastic heat engine performance (see Fig. (2)).

\section{Free: Brownian and Brownian yet non-Gaussian diffusion}
\begin{figure}
\includegraphics[scale=1.2]{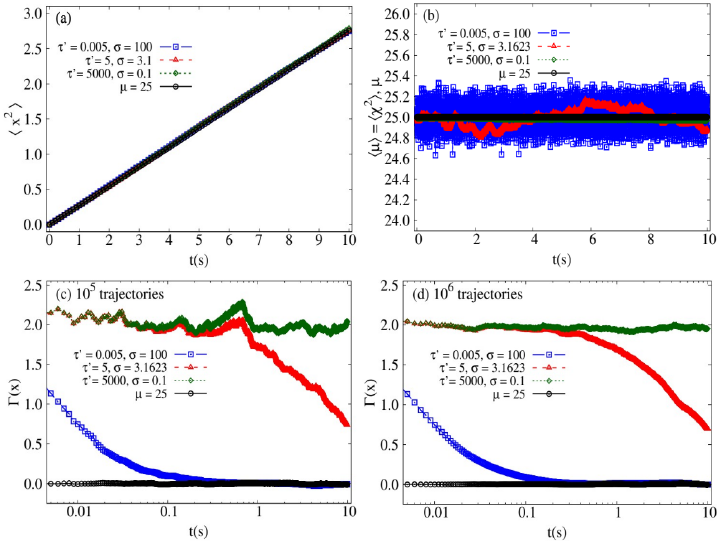}
\vspace{-0.6cm}
\caption{(a) The mean-squared displacement, (b) average mobility $\langle \mu\rangle=\langle \chi ^{2}\rangle$, (c) non-Gaussian parameter for $x$ with $10^{5}$ trajectories, and (d) non-Gaussian parameter for $x$ with $10^{6}$ trajectories are plotted as a function of time. The solid black line for Brownian diffusion. The blue, red, and green curves for Brownian yet non-Gaussian diffusion with the different sets of $\tau'$ and $\sigma$ as indicated.}
\label{fig:msd}
\end{figure}

To analyze the free diffusion in heterogeneous and homogeneous thermal baths, we substitute $\lambda(t)=0$ in Eqs. (5), and (13), respectively, of the main paper. Therefore, our Eqs. (6)-(8) becomes Eqs. (19a)-(19c) of Ref. \cite{che021}. We numerically calculated the mean-squared displacement (Fig. (\ref{fig:msd}a)), mean mobility (Fig. (\ref{fig:msd}b)), non-Gaussian parameter $\Gamma(x)$  (Fig. (\ref{fig:msd}c) with $10^{5}$ trajectories), and $\Gamma(x)$  (Fig. (\ref{fig:msd}d) with $10^{6}$ trajectories) are plotted as a function of time. Fig. (\ref{fig:msd}c) shows that $\Gamma(x)$ fluctuates highly with time. Fig. (\ref{fig:msd}a)) shows that the Brownian diffusion and Brownian yet non-Gaussian diffusion (with three sets of $\tau'$ and $\sigma$) are obeying the normal diffusion condition $\langle x^{2}(t) \rangle \propto t$. Fig. (\ref{fig:msd}b) shows average mobility fluctuations mainly depend on the value of $\sigma$. To show we need a very high number of trajectories for Brownian yet non-Gaussian diffusion \cite{che021}, $\Gamma(x)$ with $10^{6}$ trajectories is plotted in Fig. (\ref{fig:msd}d). The high fluctuation of $\Gamma(x)$ in Fig. (\ref{fig:msd}c) is reduced in Fig. (\ref{fig:msd}d). However, it is not feasible with our computing facility to consider the $10^{6}$ trajectories for the Brownian yet non-Gaussian heat engine which takes an enormous time to reach the time-periodic steady-state. Therefore, we considered only $10^{5}$ trajectories for both BHE and BNGHE for the qualitative study.

\begin{figure}
	\includegraphics[scale=1.2]{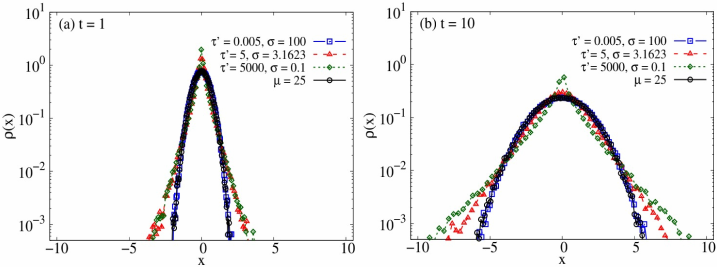}
	\vspace{-0.6cm}
	\caption{The probability distribution of $x$ is plotted at different times, (a) $t=1$, and (b) $t=10$. The solid black line for Brownian diffusion with $\mu=25$. The blue, red, and green curves for Brownian yet non-Gaussian diffusion with the different sets of $\tau'$, and $\sigma$ as indicated with $\langle\mu \rangle=25$. Y-axis in logarithmic scale.}
	\label{fig:freeprob}
\end{figure}

The probability distributions of $x$ are plotted at time, $t=1$, and $t=10$ in Fig. (\ref{fig:freeprob}), and it can be compared with the Fig. (1) of Ref. \cite{spo117}. It shows that the Brownian yet non-Gaussian diffusion consist of fast and slow-moving Brownian particles when we compare with the normal Brownian diffusion. At the bottom, the probability distributions away from the Gaussian (black curve) are due to the fast-moving Brownian particles. At the top, probability distributions above the Gaussian curves are due to the slow-moving Brownian particles as mentioned in Ref. \cite{spo117}. 

